\documentclass[reprint,aps,prc,twocolumn,superscriptaddress,floatfix,amsmath,amssymb,10pt]{revtex4-2}
\DeclareMathAlphabet{\mathpzc}{OT1}{pzc}{m}{it}
\usepackage{bm}
\usepackage{dcolumn}
\usepackage{amsthm}
\usepackage{amsmath}
\usepackage{amssymb}
\usepackage{graphicx}
\usepackage{braket}
\usepackage{xcolor}
\usepackage[mathscr]{euscript}
\usepackage{fix-cm}
\usepackage{mathptmx} 
\usepackage[T1]{fontenc}
\usepackage[colorlinks,allcolors=blue]{hyperref}
\makeatletter
\def\NAT@def@citea{\def\@citea{\NAT@separator}}
\makeatother

\begin{document}

\title{Muon-induced fission as a probe of the underlying dynamics in nuclear fission}

\author{Christian Ross}\email{christian.ross@vanderbilt.edu}
\author{A.S. Umar}\email{sait.a.umar@vanderbilt.edu}
\affiliation{Department of Physics and Astronomy, Vanderbilt University, Nashville, TN, 37235, USA}

\date{\today}

\begin{abstract}
Muon-induced fission could be utilized as a probe to study the underlying
dynamics of nuclear fission. The probability of muon attachment to the light
asymmetric fission fragment is sensitive to fission dynamics, such as the timescale
and friction of the fission event, charge asymmetry, and possibly the shape of the fission fragments.
We focus on muonic atoms that are formed with actinide nuclei.
A relativistic approach is employed, solving the Dirac equation for the muonic wavefunction in the
presence of a time-dependent electromagnetic field generated by the fissioning
nucleus. Computations are carried out on a 3-D Cartesian lattice with no
symmetry assumptions. The results show a strong dependence of the attachment probability
on the fission charge asymmetry and a more modest dependence on friction.
\end{abstract}

\maketitle

Nuclear physics experiments with muon beams provide information on fundamental symmetries and interactions.
Muonic atoms, in particular, have proven extremely useful in examining the electromagnetic properties of nuclei, e.g. electric
charge distributions and multipole moments, because the muon has a high position probability density inside the nucleus
owing to its small Compton wavelength of 1.87~fm.
Muonic atoms have been extensively utilized to study nuclear charge radii and
quadrupole moments of most stable nuclei~\cite{wohlfahrt1981,measday2001,caparthy2004,angeli2013,knecht2020,saito2025},
as well as isotope production through muon capture~\cite{ciccarelli2020,niikura2024,saito2025}.
With recent experimental advances, these studies can now be potentially extended
to neutron-rich nuclei. For heavy and neutron-rich nuclei, the muon capture
process may be delayed because most neutron levels are occupied.
When an actinide nucleus acquires a muon in an excited orbital, the resulting
excited muonic atom is also an excellent tool that may be utilized to probe the
dynamics of fission. This is largely due to the
second-generation lepton possessing a mean lifetime which exceeds typical
fission timescales by several orders of magnitude.
Muon-induced fission enjoys a rich history of interest from
both the experimental community~\cite{budick1970,chultem1975,wilcke1978,schroeder1979,ganzorig1980,johansson1980,wilcke1980,risse1991}
and the theoretical community~\cite{hadermann1976,maruhn1980,blin1982,oberacker1992,karpeshin2004}.
However, most theoretical approaches were either carried out in a non-relativistic framework
or employed only crude fission potential models in their studies.
With the advent of theoretical static and dynamic fission calculations~\cite{schunck2016,bender2020} from a microscopic
framework, it is desirable to incorporate these into the study of muon-induced fission.

\begin{figure}[!htb]
    \includegraphics*[width=8.0cm]{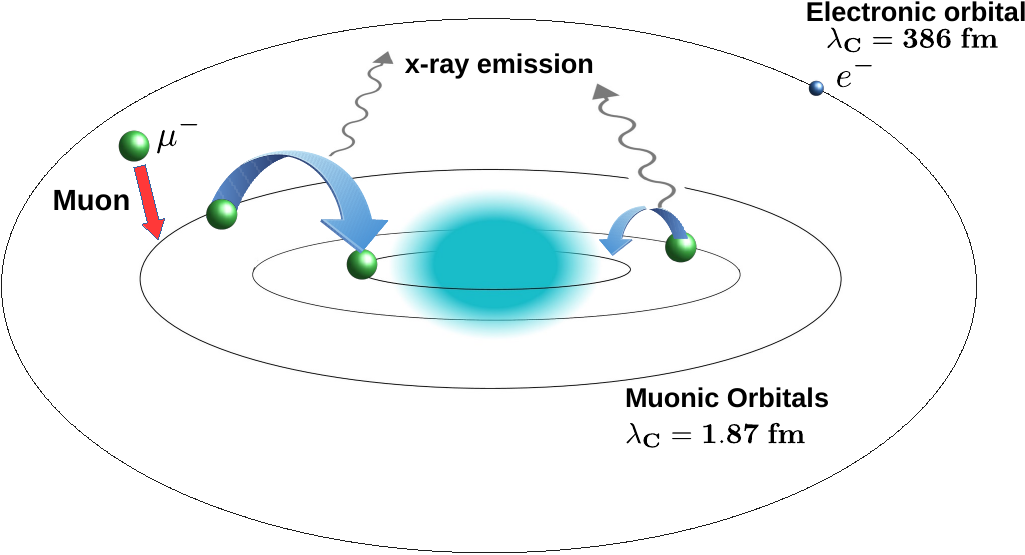}
    \caption{Depiction of the muon de-excitation process.}
    \label{fig:muon_excite}
\end{figure}

\begin{figure*}[!htb]
    \includegraphics*[width=14cm]{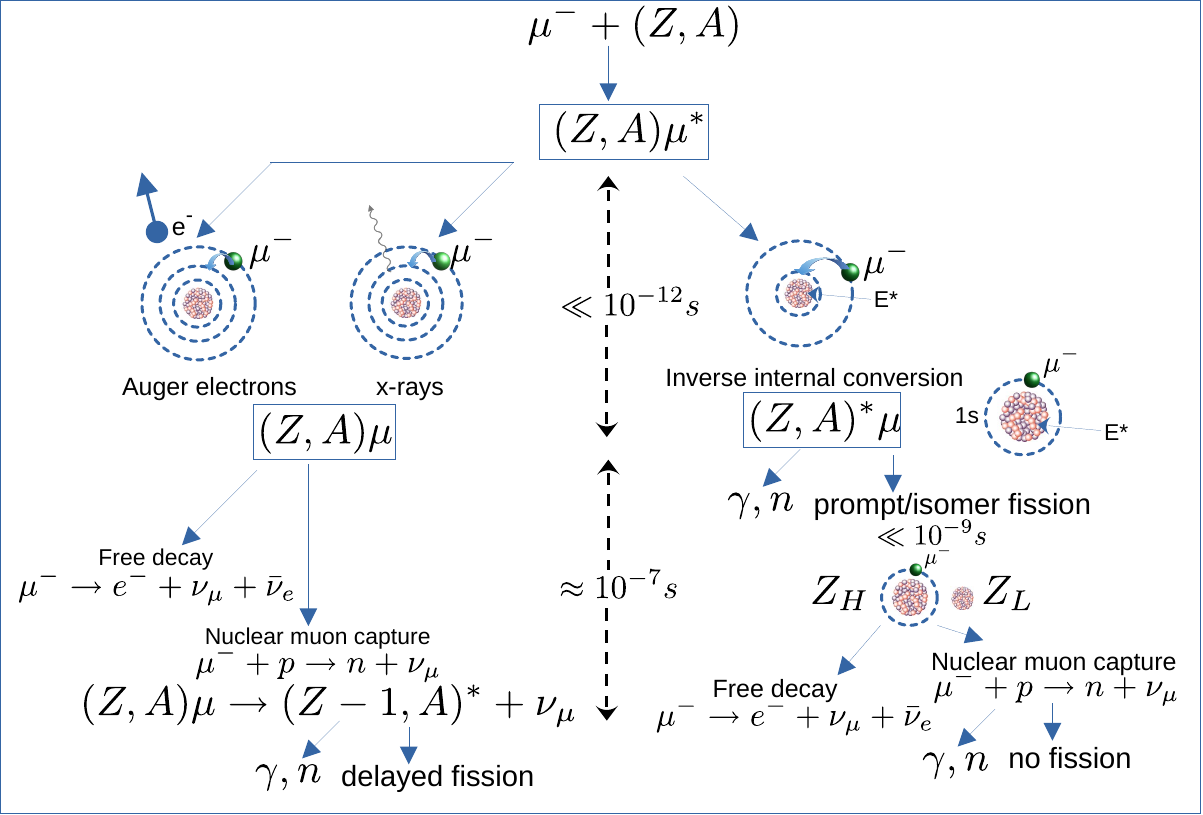}
    \caption{Possible paths a muonic atom can follow. For detailed description please see the text.}
    \label{fig:muon_capture_general}
\end{figure*}

After the formation of an excited muonic atom, the muon cascade of de-excitations begins~\cite{borie1982},
as depicted in Fig.~\ref{fig:muon_excite}.
At higher orbits ($n\approx14$) the dense level densities could lead to the ejection
of Auger electrons, while lower transitions are characterized by x-ray emission.
It is important to note the possibility of free decay of the muon
\begin{equation}
    \label{muon3}
    \mu^- \longrightarrow e^- + \nu_\mu + \overline{\nu}_e
\end{equation}
at any point of these processes.
Alternatively, it is feasible for the muon to be captured by a proton inside the
nucleus
\begin{equation}
    \label{muon2}
    (Z,A) + (\mu^-) \longrightarrow (Z-1,A)^* + \nu_\mu
\end{equation}
which leaves behind a neutron and a muon neutrino. Most of the energy is carried away by the neutrino.
The system can de-excite by emitting neutrons and gamma-rays or fissioning during this process.
Fission induced in this case is referred to as \textit{delayed} since it occurs on timescales characteristic of the weak decay
process' mean lifetime (which occurs on the order $t \approx 10^{-7}$s). In contrast, prompt fission timescales are shorter than
$t\approx  10^{-12}$s.
In Fig.~\ref{fig:muon_capture_general} we outline the most common paths
an excited muonic atom can take~\cite{wilcke1978,measday2001,adamczak2023}.

Alternatively, as the muon cascades down, inner shell transitions can occur without photon emission through inverse internal
conversion
\begin{equation}
    \label{muon1}
    (Z,A) + (\mu^-)^* \longrightarrow (Z,A)^* + (\mu^-)\;,
\end{equation}
in which the transition energy of the muon is directly transferred to the nucleus.
In actinide nuclei, the $E1 (2p\rightarrow 1s$, 6.6~MeV) and $E2 (3d\rightarrow 1s$, 9.5 MeV) muonic transitions can excite the nuclear giant dipole and giant quadrupole resonances, respectively. These resonances serve as doorway states that facilitate
the onset of fission.
An inverse internal conversion process also has the potential to excite
the nucleus into a metastable isomeric state, leading to isomeric fission.
The probability of the muon attaching to the light fission fragment is sensitive to the amount of nuclear friction experienced between the outer fission barrier and the scission point. If friction is high during the evolution from the outer barrier to scission, the muon tends to remain in the lowest molecular orbital, the $1s\sigma$ state, and consequently emerges bound in the $1s$ state of the heavy fission fragment. Conversely, if nuclear friction is small, allowing for a relatively rapid collective motion, there is a finite probability that the muon may be excited to higher-lying molecular orbitals, such as the $2p\sigma$ state, from which it can ultimately become attached to the light fission fragment. Therefore, theoretical predictions of the probability of attachment of muon to the light fragment, when combined with experimental observations, provide valuable insight into the fission dynamics and, in particular, the mechanisms of nuclear energy dissipation.
Other possibilities include back-tunneling, which describes the process of tunneling back from an isomeric state excitation rather than tunneling out to fission and also results in delayed fission.
Experimentally, when data are collected by observing muonic x-rays throughout the fission process, prompt and delayed fission modes are distinguished due to their different timescales~\cite{schroeder1979}.
Ultimately, the muon would either undergo free-decay or will be captured by one of the fission
fragments.

Since the nuclear excitation energy in muon-induced fission is larger than the height of the fission barrier, the fission dynamics can be reasonably treated within a classical framework, without the need to consider quantum tunneling through the barrier. For simplicity, the evolution of the fission process is described using a single collective coordinate,
$R$, representing the separation between the fragments. In this classical picture, the collective nuclear energy takes the form:
\begin{equation}
    E_{nuc} = \frac 12 B(R)\dot R^2 + V_{fis}(R) + E_{\mu} (R)
    \label{eq:enuc}
\end{equation}
where a coordinate-dependent mass $B(R)$ is introduced~\cite{oberacker1993}.
As the nucleus moves from the outer
fission barrier to the scission point one can also introduce a dissipation term, which depends quadratically on the relative
velocity
\begin{equation}
    \frac{dE_{nuc}}{dt} = -\frac{dE_{diss}}{dt}=-f\dot R^2(t).
    \label{eq:dynamics}
\end{equation}
This constant parameter, $f$, can be determined from the time-dependent Hartree-Fock (TDHF) evolution~\cite{iwata2022}, but in general can be time-dependent~\cite{qiang2021b}.
Equation~\eqref{eq:dynamics} together with Eq.~\eqref{eq:enuc} determines the equation
for the collective dynamics to be solved, subject to the initial condition that the
kinetic energy contains the excitation energy, $E^*=9.5$~MeV, corresponding to the
$3d\rightarrow 1s$ transition with inverse internal conversion.

The calculation of the fission potential, $V_{fis}(R)$, proceeds in two steps,
we first perform constrained Hartree-Fock (CHF) calculations of the fission potential by
constraining the quadrupole moment of the nucleus in a standard manner, described
for our case in Ref.~\cite{godbey2024}. Starting from a point after the saddle
point we evolve the system using TDHF equations~\cite{simenel2018}
to a large $R$ value (about 80-100~fm),
During this evolution we use the density constrained time-dependent Hartree-Fock (DC-TDHF) calculations
to obtain the $V_{fis}(R)$ values.

In describing the dynamical evolution of the muonic wavefunction during prompt fission, the dominant interaction arises from the electromagnetic coupling between the muon and the nucleus, characterized by the term $-e\gamma_{\mu}A^{\mu}$.
The contribution from the weak interaction is negligible in this context. Furthermore, since the motion of the fission fragments is
nonrelativistic, the electromagnetic interaction is primarily governed by the Coulomb force.
The muon is a relativistic entity (in the ground state of an actinide muonic
atom, the muonic binding energy is approximately 12$\%$ of the muonic rest mass), which
strongly implies that calculations performed with the Schr\"{o}dinger equation
are quite limited (note that some groups have shown that this is feasible~\cite{oberacker2004}).
Instead, we solve the Dirac equation for the muon in the
presence of a time-dependent electromagnetic field generated by the fissioning
nucleus.
The Dirac equation for a muonic spinor wavefunction in the presence of an
external electromagnetic field is given through the minimal coupling
prescription
\begin{equation}
\label{diraceq1}
\bigg (i\hbar c \gamma^{\mu}\partial_{\mu} - q\gamma^\mu A_\mu (\mathbf r, t)- mc^2 \bigg)\psi(\mathbf r, t) = 0\;.
\end{equation}
Assuming that the motion of the fission fragments is nonrelativistic, the
Coulomb interaction will dominate. The scalar Coulomb potential $A_0(\mathbf r,
t)$ is several orders of magnitude larger than the vector potential
$\mathbf A(\mathbf r, t)$ components of the electromagnetic four-potential so the vector
potential is neglected. The Dirac Hamiltonian can be written as
\begin{equation}
    \label{dirachamiltonian1}
    \hat H_D = -i\hbar c \alpha^i \partial_i + \beta mc^2 - qA_0(\mathbf r, t)\;,
\end{equation}
with the scalar Coulomb potential obtained by
using the proton densities from CHF and TDHF calculations and employing fast Fourier transform (FFT) techniques for the Poisson equation
\begin{equation}
    \label{coulombpot1}
    \nabla^2 A_0(\mathbf r, t) = -4\pi q^2 \rho_p (\mathbf r, t)\;.
\end{equation}
The solution of the Dirac equation proceeds in two steps: first, a static
solution is generated by solving the stationary Dirac equation for the
ground-state spinor
\begin{equation}
    \label{diraceq3}
    \hat H_D(\mathbf r, 0) \psi_{gs}(\mathbf r) = E_{gs}\psi_{gs}(\mathbf r)\;.
\end{equation}
Once a static ground-state solution is found, the time-dependent Dirac equation
\begin{equation}
    \label{diraceq2}
    \hat H_D \psi(\mathbf r, t) = i\hbar \frac{\partial}{\partial t} \psi(\mathbf r, t)\;,
\end{equation}
is propagated in time via a Taylor expansion of the unitary time-evolution operator
\begin{equation}
    \label{diraceq5}
    \psi(t+\Delta t) \approx \bigg(\mathbb I + \sum_{n=1}^N \frac{(-i\Delta t \hat H_D)^n}{n!} \bigg) \psi(t)\;,
\end{equation}
with $\psi(\mathbf r, t=0) =  \psi_{gs}(\mathbf r)$.
The instantaneous muonic binding energy $E_{\mu}$, to be used in Eqs~\eqref{eq:enuc} and \eqref{eq:dynamics}, is given by
\begin{equation}
    E_{\mu}(R(t)) = \braket{\psi(\mathbf r, t) | \hat H_D|\psi(\mathbf r, t)}\;.
    \label{eq:emu}
\end{equation}
\begin{figure*}[htb!]
    \includegraphics*[width=16cm]{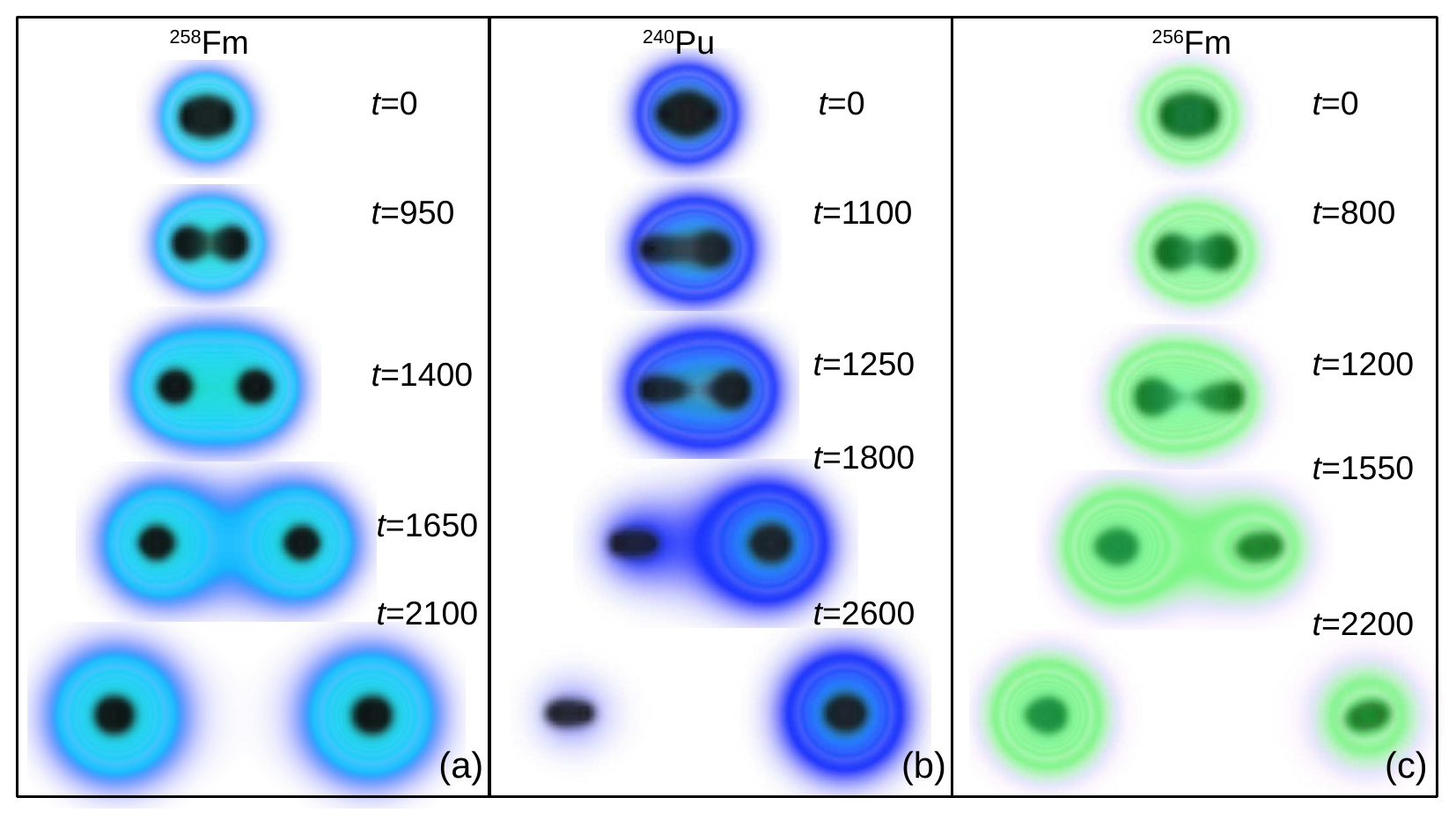}
    \caption{Time-evolution of the muon density (a) $^{258}$Fm: charge asymmetry of $1.0$, (b)
        $^{240}$Pu: charge asymmetry of $1.29$, (c) $^{256}$Fm: charge asymmetry of $1.15$. 
        Times are given in units of fm/c. The dark shaded
        region depicts the nuclear fragment densities.\label{fig:time-evolv}}
\end{figure*}

We have written a new code to solve the Dirac equation
on a three-dimensional Cartesian lattice with no
symmetry assumptions using the basis-spline discretization~\cite{umar1991b}
method. The static muon wave functions are obtained using the relativistic
version of the gradient iteration method~\cite{bottcher1989}.
For the three systems studied here, $^{240}$Pu, $^{256,258}$Fm, we vary the quadrupole moment from the ground-state value
up to $7400-8000$~fm$^2$ in steps of 100~fm$^2$ and a CHF with Bardeen-Cooper-
Schrieffer (BCS) pairing
calculation is performed at each constraint step, using the SLy4d Skyrme interaction~\cite{kim1997}, with a density dependent delta interaction (DDDI) in the pairing channel using neutron and proton strengths 1256~MeV~fm$^3$ and 1462~MeV~fm$^3$, respectively.
The fission potential for $^{240}$Pu is doubled humped as shown in Ref.~\cite{godbey2024},
with an inner barrier height of 6.0~MeV and outer barrier height of 3.9~MeV.
Whereas the $^{256,258}$Fm potentials are single humped~\cite{scamps2015a},
with barrier heights of 6.0 and 5.5~MeV, respectively.
Past this point, the DC-TDHF dynamic run evolves the system through the
remainder of the fission process.
CHF and DC-TDHF calculations were performed on a numerical mesh of size $140 \times 32\times 32$~fm$^3$.
The proton densities are then placed on a larger mesh of size
$140\times70\times70$~fm$^3$ for the muon fission code. This is required since the
muonic wavefunction spreads out over a relatively large volume.
The muon attachment probability is determined by integrating the muon density
to the left and right halves of the 3D numerical box, as
\begin{equation}
    \label{muondensity1}
    \begin{split}
        P_L(t) =  &\int d^3x \; \psi^{\dagger}(\mathbf x, t)\psi(\mathbf x, t) \;\; ; \;\; x < 0 \\
        P_R(t) =  &\int d^3x \; \psi^{\dagger}(\mathbf x, t)\psi(\mathbf x, t) \;\; ; \;\; x > 0\;,
    \end{split}
\end{equation}
satisfying $P_L+P_R=1$.

As we have mentioned above, the muon attachment probability to the fission fragments is sensitive to fission
dynamics and the charge asymmetry of the fragments.
$^{240}$Pu fissions with a charge asymmetry of $Z_H/Z_L=1.29$.
$^{258}$~Fm and $^{256}$~Fm have bimodal fission modes, for $^{258}$~Fm the symmetric fission mode is more favorable, while for $^{256}$~Fm the asymmetric mode appears to be favorable~\cite{staszczak2007}.
For $^{258}$~Fm we chose the
symmetric fission path, and for $^{256}$Fm we chose a path with charge asymmetry of 1.15,
corresponding to the heavy fragment having $Z=54$.
Sometimes the additional contribution from muon energy in Eq.~(\ref{eq:enuc}) has been interpreted as modifying the original fission potential, in our case
the modification of the fission barriers due to the muon energy for $^{240}$Pu are 0.19~MeV and
0.98~MeV at the inner and outer barriers, respectively. For  $^{256}$Fm and $^{258}$Fm nuclei
these modifications are 0.24~MeV and 0.33~MeV, respectively.
\begin{table}[h]
    \caption{Results for  $^{240}$Pu, $^{256}$Fm, and $^{258}$Fm systems. For details see text.}
    \begin{ruledtabular}
    \begin{tabular}{lccc}
        System & $^{240}$Pu & $^{256}$Fm & $^{258}$Fm \\ \hline
        Charge asymmetry & 1.29   & 1.15   & 1.0    \\
        Mass asymmetry & 1.30   & 1.16   & 1.0    \\
        $P_L$ (no friction) & 5.2\% & 20.6\% & 50.0\% \\
        f (MeV/fm$\cdot$c) & 456 & 550 & 562 \\
        $E_{diss}$ (MeV) & 21.94 & 29.60 & 18.41 \\
        $P_L$ (friction) & 6.5\% & 16.6\% & 50.0\%
    \end{tabular}
    \end{ruledtabular}
    \label{table:muon}
\end{table}

In Table~\ref{table:muon} we show the summary of results at the final asymptotic time-step for the three systems. We
note that charge asymmetries are about the same as their mass asymmetries, $A_H/A_L$, for these systems.
We see a strong dependence of the probability of attachment of muons
to the light fragment, $P_L$, on this asymmetry. For the plutonium case with a mass
asymmetry of 1.30, this probability is 5.2\%, whereas for the $^{256}$Fm case with
asymmetry value of 1.16 the probability is 20.6\%. For the symmetric fission of
$^{258}$Fm, the muonic wavefunction is evenly split,
and the probability is exactly 50\%, which also serves as a computational
test case. For these cases, when friction is included, these probabilities change
to 6.5\%, 16.6\%, and again 50\%. 
The increase of the light fragment attachment probability in the case of $^{240}$Pu from
     5.2\% to 6.5\% does not fit the expectation that with increasing friction this probability
     should decrease. We have further tested this by artificially increasing the friction
     parameter and observed that the probability slowly decreases to 0\% in the adiabatic limit.
     Thus we interpret this being a result of the sensitivity of the model to the
     constant friction model parameter f.
The trend for the dissipated energy is similar as well,
with the results being commensurate with those given in~\cite{iwata2022}.
Figure~\ref{fig:time-evolv} shows the time-evolution snapshots of the muon density for (a) $^{258}$Fm, (b)
$^{240}$Pu, and (c) $^{256}$Fm, starting from the ground state to the final separation distance of
about 100~fm.
The dark shaded region depicts the nuclear fragment densities.

Employing a muon as a probe of the underlying dynamics involved in nuclear
fission is a promising tool. Due to the sensitivity of the muon attachment
probability to the fission fragments, insight can be gained into several
physical mechanisms, such as the nuclear dissipation energy and the fission timescales.
We have written a new code to solve the Dirac equation for a muonic spinor wavefunction in the presence of an
external time-dependent electromagnetic field generated by the fissioning
nucleus. CHF and DC-TDHF calculations were used to generate
fission potentials, proton densities, and Coulomb potentials for
fissioning systems, $^{240}$Pu, $^{256}$Fm, and $^{258}$Fm. The results show a strong
dependence of the muon attachment probability to the light fragment on the fragment charge
and mass asymmetry. The dependence on friction, as was modeled in our study, was more modest.
There is still the question whether the muon prefers the spherical fragment
more compared to the deformed one, which requires more systematic studies.
Our long-term goal is to interface our muon code with more sophisticated
fission codes and models. In principle, the collective motion of the fissioning nucleus
can be integrated with dynamical microscopic fission calculations as was done
in Ref.~\cite{qiang2021b,qiang2021}, which also allows for a time-dependent friction calculation.

\begin{acknowledgments}
This work has been supported by the U.S. Department of Energy under award number DE-SC0013847 (Vanderbilt University).
\end{acknowledgments}

\bibliography{VU_bibtex_master}

\end{document}